\documentclass[aps,prl,unsortedaddress,superscriptaddress,twocolumn]{revtex4}
\usepackage{amsmath}
\usepackage{amssymb}
\usepackage{bm}
\usepackage{epsfig}
\usepackage{graphicx}

\input{tcilatex}

\begin{document}

\title{Qubits based on Polariton Rabi Oscillators}
\author{S.S. Demirchyan}
\affiliation{Department of Physics and Applied Mathematics, Vladimir State University
named after A.G. and N.G. Stoletovs, Vladimir, 600000, Russia}
\author{I.Yu. Chestnov}
\affiliation{Department of Physics and Applied Mathematics, Vladimir State University
named after A.G. and N.G. Stoletovs, Vladimir, 600000, Russia}
\author{A.P. Alodjants}
\affiliation{Department of Physics and Applied Mathematics, Vladimir State University
named after A.G. and N.G. Stoletovs, Vladimir, 600000, Russia}
\affiliation{Russian Quantum Center, Novaya str. 100, 143025, Skolkovo, Moscow, Russia}
\author{M.M. Glazov}
\affiliation{Ioffe Physical-Technical Institute of the RAS, 26 Polytekhnicheskaya,
St.-Petersburg 194021, Russia}
\affiliation{Spin Optics Laboratory, St. Petersburg State University, 1 Ul'anovskaya,
Peterhof, St. Petersburg 198504, Russia}
\author{A.V. Kavokin}
\affiliation{Spin Optics Laboratory, St. Petersburg State University, 1 Ul'anovskaya,
Peterhof, St. Petersburg 198504, Russia}
\affiliation{School of Physics and Astronomy, University of Southampton, SO17 1NJ,
Southampton, UK}

\begin{abstract}
We propose a novel physical mechanism for creation of long lived macroscopic
exciton-photon qubits in semiconductor microcavities with embedded quantum
wells in the strong coupling regime. The polariton qubit is a superposition
of lower branch (LP) and upper branch (UP) exciton-polariton states. We
argue that the coherence time of Rabi oscillations can be dramatically
enhanced due to their stimulated pumping from a permanent thermal reservoir
of polaritons. We discuss applications of such qubits for quantum
information processing, cloning and storage purposes.
\end{abstract}

\maketitle





\textit{Introduction}. --- Polaritonics is an interdisciplinary research
area at the boundary of optics and solid state physics. It is aimed at the
studies of light-matter interaction and dynamics of exciton-polaritons, or
shortly polaritons: quasi-particles with bosonic statistics formed as a
result of the light-exciton coupling. Nowadays polaritonics represents an
indispensable tool for investigation of quantum coherent and nonlinear
phenomena occurring at the matter-field interface in various area of
condensed matter physics, quantum and atom optics~\cite%
{Annu.Rev.Matter.37.317,Phys.Rev.A.65.022314,Nat.Phys.2.849,Phys.Rev.A.84.013813}%
.

Semiconductor microcavities serve as a solid-state laboratory to study
dynamical and quantum effects in open and non-equilibrium systems of bosons.
Particularly, one of the main achievements in the field of polaritonics is
the creation of and manipulation with condensates characterized by a
macroscopic occupation of a single quantum state and extended temporal and
spatial coherences. In this sense, polaritonics presents a significant
interest for quantum information science.

Recently, various approaches have been proposed for classical and quantum
computation with use of microcavity polaritons, see e.g.~\cite%
{PRL.99.196402, NatureComms.4.1778, PRB.69.245304, PRA.86.052313,
J.Rus.Las.Res.27.482}. It has been proposed in \cite{PRL.99.196402} and then
demonstrated in~\cite{NatureComms.4.1778} that the classical information can
be carried by lower branch (LP) polaritons propagating in microcavity based
optical integrated circuits. In quantum domain, an approach for generation
of branch-entangled pairs of polaritons in microcavities with use of the
spontaneous interbranch parametric scattering has been formulated \cite%
{PRB.69.245304} opening up a possibility to use this process for quantum
information processing~\cite{PRA.86.052313}.

The main advantage of using exciton-polaritons for quantum information
processing purposes comes from their fast switching properties (the typical
switching time of a few picoseconds), relatively strong nonlinear response,
low power to perform logical operations \cite{NatureComms.4.1778} and
features of superfluid propagation, which are {essential} for realization of
many algorithms in quantum information science, see e.g. \cite%
{Nielsen_quant_comp, Nature:154.45.2010}. Current polaritonic devices could
be designed by using very well developed semiconductor micro- and
nanotechnologies and enable to operate at high temperatures, up to the room
temperature \cite{Christopolous}. However, due to open and non-equilibrium
nature of the polariton condensates, they cannot serve directly as qubits,
since it is not possible so far to generate a polariton state with a fixed
well-defined number of particles.

In practice, macroscopic polaritonic system could be used for continuous
variable quantum computation, or for quantum comutation with macroscopic
polariton states, cf.~\cite{PRA.61.042309,PRA:86.023829}. Light-matter
interaction in microcavities gives rise to the natural two-level system, a
Rabi doublet, or a doublet of lower, $|LP\rangle $, and upper, $|UP\rangle $%
, \emph{macroscopically} occupied orthogonal polariton states 
\begin{equation}
|LP\rangle =C_{x}|X\rangle -C_{p}|P\rangle ,\quad |UP\rangle =C_{p}|X\rangle
+C_{x}|P\rangle ,  \label{doublet}
\end{equation}%
being hybridized states of the quantum well exciton, $|X\rangle $, and
cavity photon, $|P\rangle $. Hopfield coefficients $C_{x,p}=2^{-1/2}(1\pm {%
\Delta }/\sqrt{4g^{2}+\Delta ^{2}})^{1/2}$ are determined by the system
parameters with $g$ being the exciton-photon coupling parameter and $\Delta $
being the detuning between the bare photon and exciton mode and can be
controlled in the state-of-the-art structures with needed accuracy. The
quantum state of a qubit $\left\vert \Psi \right\rangle $ can be presented
as a linear combination of $|LP\rangle $ and $|UP\rangle $ states, 
\begin{equation}
\left\vert \Psi \right\rangle =\beta _{1}|UP\rangle +\beta _{2}|LP\rangle ,
\label{qubit}
\end{equation}%
with two complex coefficients $\beta _{1,2}$ satisfying the normalization
condition $|\beta _{1}|^{2}+|\beta _{2}|^{2}=1$. Hence, the quantum state of
a qubit is determined by the occupations of the upper and lower polariton
states given by $|\beta _{1,2}|^{2}$, respectively, as well as by their
relative phase. In the case of free evolution of the system the coefficients
in Eq.~\eqref{doublet} read: $\beta _{1,2}=e^{-i\Omega _{1,2}t}\left/ \sqrt{2%
}\right. $, where $\Omega _{1,2}$ are the eigenfrequencies of UP and LP
states, respectively. The beats between LP and UP states are known as Rabi
oscillations.

Current progress in the microcavity growth technology makes it possible to
produce structures where such Rabi-oscillator based qubits can be coupled to
each other thus paving way to polariton-based quantum computation devices.
However, the main problem here is to preserve a coherence between $%
|UP\rangle $ and $|LP\rangle $ states within the time of quantum logic
operations, or, most generally, during the time of computation \cite%
{Nielsen_quant_comp}. Actually, decoherence occurring due to the interaction
of the qubit system with its environment prevents application of quantum
algorithms \cite{Proc.R.Soc.Lond.A.452.567}. {As a result,} in the presence
of decoherence a polariton\ qubit state \eqref{qubit} decays as $\left\vert {%
\Psi }\right\rangle \propto e^{-t\left/ \tau _{R}\right. }$, where $\tau
_{R} $ is characteristic decay time {governed by the scattering of
polaritons to the reservoir and the photon leakage through the mirrors}. The
decoherence time may be quite short in realistic systems, typically
comparable or even shorter than the lifetime of exciton-polariton, $\tau
_{0} $, that is on the picosecond timescale \cite{Timofeev}. Thus, only a
few periods of polariton Rabi-oscillations could have been observed
experimentally \cite{Gibbs, Masha}.

The aim of this Letter is to indicate the way of creating of a stable
polariton qubit in a resonantly \emph{cw} pumped system of
exciton-polaritons, where Rabi oscillations are induced by a short pulse of
light. We demonstrate that in the presence of an incoherent reservoir of
polaritons 
the coherence time of Rabi oscillations may be dramatically increased. This
is because of the stimulated scattering of polaritons towards the qubit
state $\left\vert \Psi \right\rangle $, which supports the given
superposition of $|LP\rangle $ and $|UP\rangle $ states. In realistic
microcavity systems the enhancement of the coherence time up to nanoseconds
can be achievable.


\textit{Model.} --- We consider exciton-polaritons in a planar or pillar
microcavity under incoherent nonresonant \emph{cw} pumping, see Fig. \ref%
{fig:scheme}. 
\begin{figure}[tbp]
\includegraphics[width=0.45\textwidth]{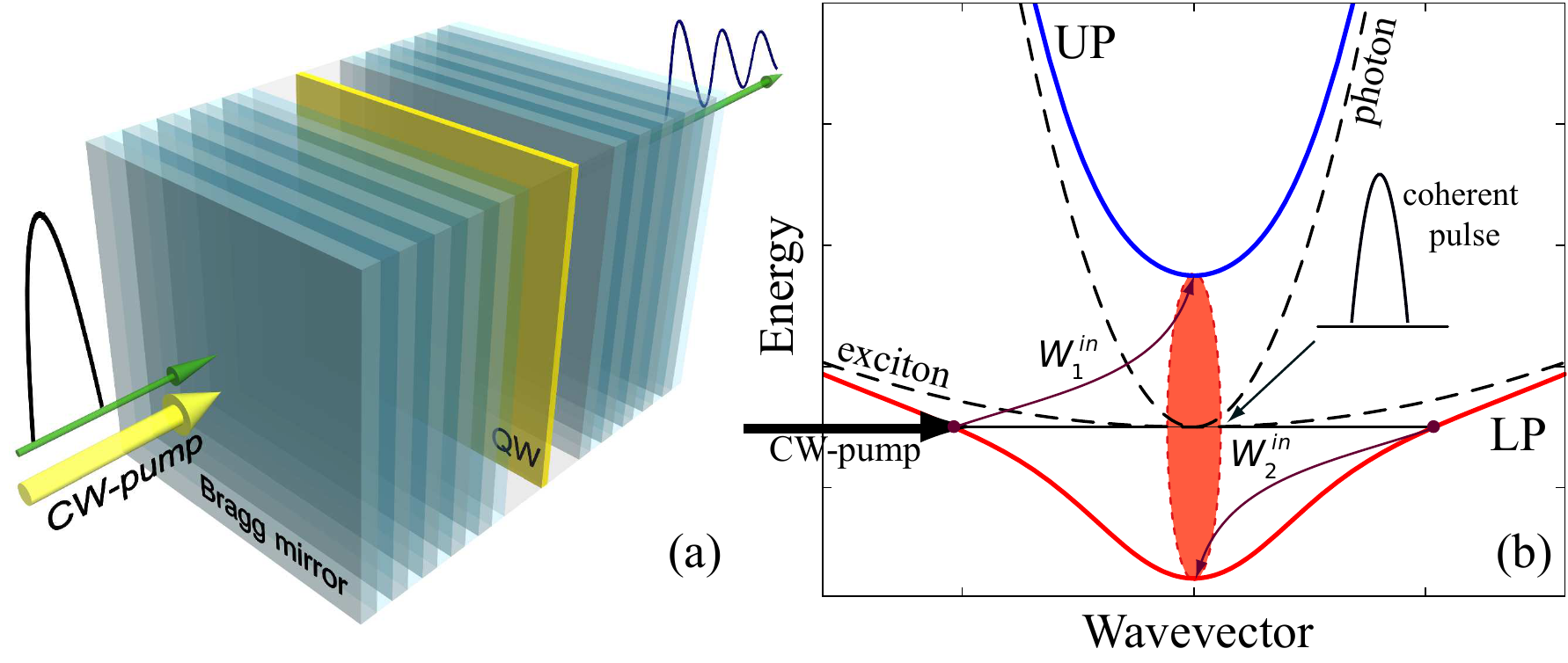}
\caption{(Color online) (a) Scheme of polariton qubit excitation in
semiconductor microcavity with embedded quantum well (QW) sample. (b)
Polariton dispersion and schematic polariton scattering processes supported
by incoherent reservoir pumping (shadow area) for $\Delta =0$.}
\label{fig:scheme}
\end{figure}
The pumping helps forming a reservoir of long-living exciton-polaritons with
large in-plane wavevectors. We assume that the pumping is strong enough so
that only two polariton states $|LP\rangle $ and $|UP\rangle $ can be
macroscopically occupied. The dynamics of the Rabi doublet can be most
conveniently described by the density matrix approach. The diagonal elements
of a $2\times 2$ density matrix $\varrho $ determine the mean occupations $%
\varrho _{11}=N_{1}$ of the $|UP\rangle $ and $\varrho _{22}=N_{2}$ of the $%
|LP\rangle $ states, while off-diagonal elements $\varrho _{12}=\varrho
_{21}^{\ast }$ determine the coherence between those states. For a pure
quantum state in a form of Eq.~\eqref{qubit} $\varrho _{11}=|\beta _{1}|^{2}$%
, $\varrho _{22}=|\beta _{2}|^{2}$, $\varrho _{12}=\beta _{1}\beta
_{2}^{\ast }$ and $\varrho _{21}=\beta _{1}^{\ast }\beta _{2}$. In this
Letter, we refrain from the analysis of the full statistics of the polariton
Rabi doublet, which can be treated by other methods~\cite%
{laussy:book,PhysRevB.80.195309}, and follow the pseudospin approach like in
Ref.~\cite{glazov_sns_pol}. For simplicity we neglect polariton-polariton
interactions for the macroscopically occupied states. The dynamics of the
density matrix is given by $\frac{d\varrho }{dt}=-\frac{\mathrm{i}}{\hbar }[%
\mathcal{H},\varrho ]+\mathcal{L}\{\varrho \},$ where $\mathcal{H}$ is the
Hamiltonian of the system whose only nonzero elements are the diagonal ones, 
$\mathcal{H}_{11}=E_{1}$, $\mathcal{H}_{22}=E_{2}$, being energies of $%
|UP\rangle $ and $|LP\rangle $ states, and $\mathcal{L}\{\varrho \}$ stands
for the Lindblad superoperator describing dissipation in the system.
Equations describing occupation dynamics have a standard form: 
\begin{equation}
\dot{\varrho}_{ii}=\dot{N_{i}}=-\frac{N_{i}}{\tau _{i}}+(1+N_{i})W_{i}^{%
\mathrm{in}}-N_{i}W_{i}^{\mathrm{out}},\ \ \ i=1,2,  \label{rho}
\end{equation}%
where dots denote time derivatives and $\tau _{i}$ is the lifetime of the $i$%
th state. In particular, $\tau _{1,2}$ are expressed through exciton ($\tau
_{exc}$) and photon ($\tau _{ph}$) lifetimes as $\tau _{1,2}^{-1}=\left\vert
C_{p,x}\right\vert ^{2}\tau _{exc}^{-1}+\left\vert {C_{x,p}}\right\vert
^{2}\tau _{ph}^{-1}$. For the state-of-the-art semiconductor microcavities
where $\tau _{exc}^{{}}\sim 100$~ps and $\tau _{ph}\sim 10$~ps the
inequality $\tau _{ph}\ll \tau _{exc}$ typically holds, cf.~\cite{Timofeev}.
In this case one can set $\tau _{1,2}\simeq \tau _{ph}^{{}}/\left\vert
C_{x,p}\right\vert ^{2}$ for LP and UP branch lifetimes, respectively. In
Eqs.~\eqref{rho}, $W_{i}^{\mathrm{in}}$ ($W_{i}^{\mathrm{out}}$) is the
in-scattering rate to (out-scattering rate from) the state $i$ from (to) the
reservoir, Fig. \ref{fig:scheme}b. In general, $W_{1}^{\mathrm{in/out}}\neq
W_{2}^{\mathrm{in/out}}$ owing to the significant Rabi splitting $\hbar
\Omega _{R}=\left\vert E_{1}-E_{2}\right\vert =\hbar \sqrt{\Delta ^{2}+4g^{2}%
}$, which can be comparable with the energy of exciton-polaritons in the
reservoir. As follows from Eqs.~\eqref{rho} any fluctuation $\delta
N_{i}=N_{i}-\bar{N}_{i}$, where $\bar{N}_{i}$ are the steady occupations of
polariton states, decays according to $\dot{\delta N}_{i}=-{{\delta N_{i}}/{%
\tau _{c,i}}},$ where $\tau _{c,i}=(\bar{N}_{i}+1)(1/\tau _{i}+W_{i}^{%
\mathrm{out}})^{-1}$, which is the longer, the larger is the occupation of
the state~\cite{glazov_sns_pol}.

\begin{figure}[t]
\includegraphics[width=0.45\textwidth]{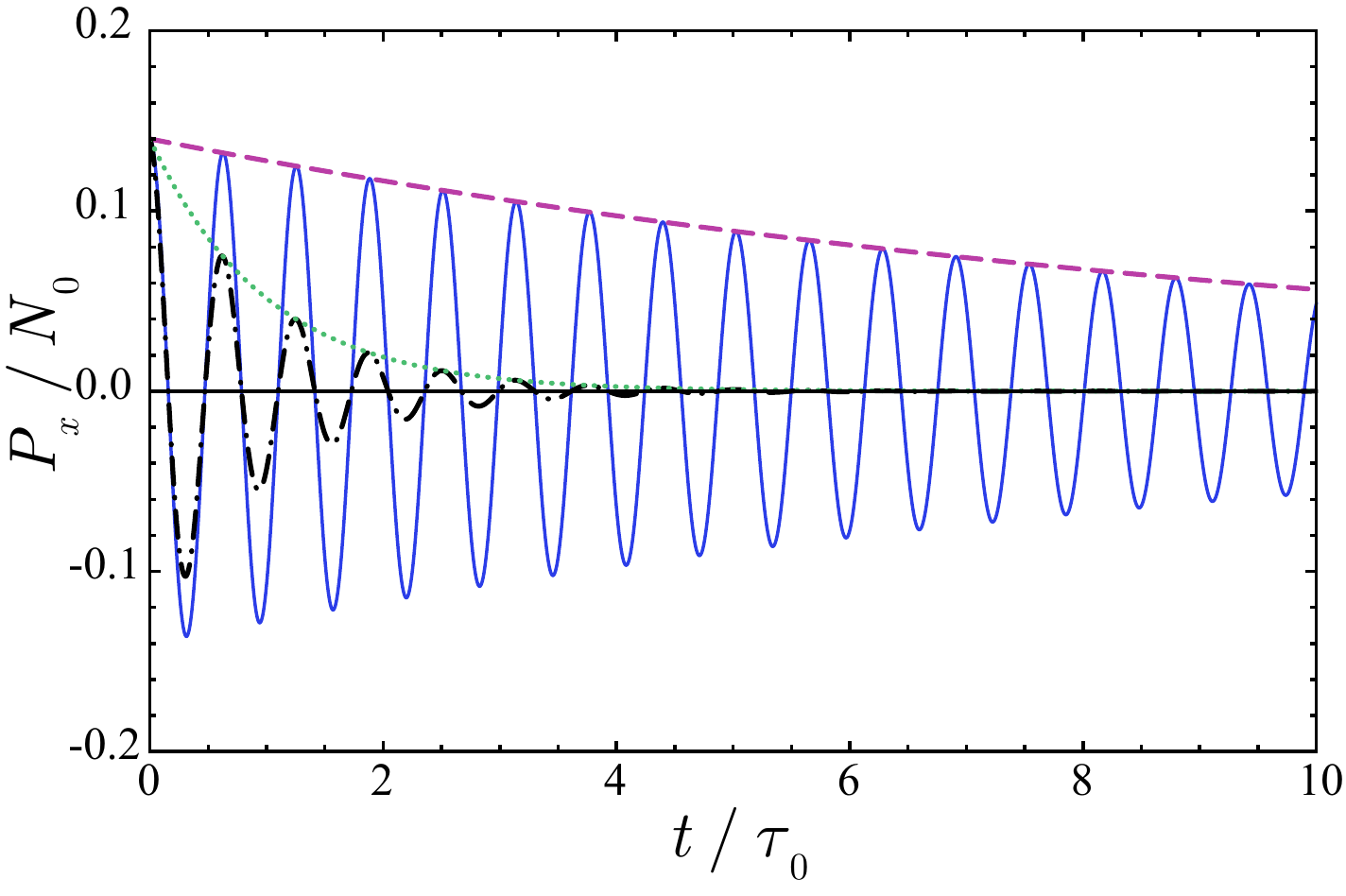}
\caption{(Color online) Temporal dynamics of normalized Bloch vector
component $P_x/N_{0}$ ($N_{0}\equiv N(t=0)$) in the presence (solid/blue
curve) and in the absence (dash-dotted/black curve) of pumping.
Dashed/purple and dotted/green curves show the envelope of Rabi oscillations
in the presence and absence of pumping, respectively. The parameters are: $P%
\protect\tau_0 = 20$, $W^{\mathrm{out}}_{1,2} =0$, $\Omega_R\protect\tau%
_0=10 $. Initial conditions at $t=0$ are: $P_{x}{/N_0}= 0.14$, $P_{y}{/N_0}%
=0 $ and $P_z {/N_0}= 0.48$.}
\label{fig:Rabi}
\end{figure}

To address the dynamics of Rabi oscillator it is convenient to parametrize
the density matrix using the pseudospin formalism: $\varrho _{11}=N+P_{z}$, $%
\varrho _{22}=N-P_{z}$, $\varrho _{12}=P_{x}-\mathrm{i}P_{y}$. Here $%
N=(N_{1}+N_{2})/2$ and $\bm P=(P_{x},P_{y},P_{z})$ is the pseudospin.
Following \cite{PhysRevLett.92.017401,glazov_sns_pol} we obtain 
\begin{subequations}
\label{dynamic_eq}
\begin{eqnarray}
{\dot{N}}{=-\left[ \tau _{+}^{-1}-\delta W_{+}\right] N-\left[ \tau
_{-}^{-1}-\delta W_{-}\right] P_{z}+W_{+}} &&,  \label{N} \\
{\dot{P}_{z}}{=-\left[ \tau _{+}^{-1}-\delta W_{+}\right] P_{z}-\left[ \tau
_{-}^{-1}-\delta W_{-}\right] N+W_{-}} &&,  \label{Pz} \\
{\dot{\bm P}_{\bot }}{=-\left[ \tau _{+}^{-1}+\tau _{\bot }^{-1}-\delta W_{+}%
\right] {\bm P}_{\bot }-\left[ {\bm\Omega _{R}}\times {\bm P}_{\bot }\right] 
} &&.  \label{Pxy}
\end{eqnarray}%
Here $\bm P_{\perp }=(P_{x},P_{y})$, $\bm\Omega _{R}=\Omega _{R}\bm e_{z}$,
where $\bm e_{z}$ is a unit vector along $z$ axis, $W_{\pm }=(W_{1}^{\mathrm{%
in}}\pm W_{2}^{\mathrm{in}})/2$, $\delta W_{\pm }=[(W_{1}^{\mathrm{in}%
}-W_{1}^{\mathrm{out}})\pm (W_{2}^{\mathrm{in}}-W_{2}^{\mathrm{out}})]/2$,
and $1/\tau _{\perp }$ is the additional damping rate for the off-diagonal
density matrix components. We stress that the pseudospin $\bm P$ is
equivalent to the Bloch vector used to describe the state of any two-level
system. In Eqs.~\eqref{dynamic_eq} we have introduced the characteristic
decay rates $\tau _{\pm }^{-1}=\left( \tau _{1}^{-1}\pm \tau
_{2}^{-1}\right) /2$ .


\textit{Results and discussion.} Figure \ref{fig:Rabi} shows the temporal
dynamics of the normalized pseudospin Bloch vector component $P_{x}$ {%
calculated numerically from} Eqs.~\eqref{dynamic_eq}  {assuming that} $\tau
_{\perp }\gg \tau _{0}$ and $C_{x,p}=1\left/ \sqrt{2}\right. $, $W_{1}^{%
\mathrm{in}}=W_{2}^{\mathrm{in}}$. 
{Other parameters of the calculation are presented in the caption to Fig.~%
\ref{fig:Rabi}.} We assumed that at $t=0$ the coherence between the upper
and lower polariton branches is established by a short and weak laser pulse
(see Fig.~\ref{fig:scheme}), that sets the initial conditions in Fig.~\ref%
{fig:Rabi}. {As it is clearly seen from Fig.~\ref{fig:Rabi}, the lifetime of
Rabi oscillations $\tau _{R}$ increases with the \emph{cw} pumping intensity.%
} In such a case, the Rabi oscillations are sustained by the reservoir.

{To evaluate this effect analytically, we note that in accordance with Eqs.~%
\eqref{dynamic_eq} ${\bm P}_{x}+i{\bm P}_{y}\propto e^{-i\Omega _{R}t-{t{%
\left/ {\vphantom {t \tau _{R} }}\right. }\tau _{R}}}$ where the decay rate
of Rabi oscillations $\tau _{R}^{-1}$ is given by}

\end{subequations}
\begin{equation}
\tau _{R}^{-1}=\tau _{0}^{-1}-\delta W_{+}.  \label{tau}
\end{equation}%
{Here $\tau _{0}=\tau _{+}\tau _{\perp }\left/ {}\right. \left( \tau
_{+}+\tau _{\perp }\right) $ is the effective lifetime of the polaritonic
system without reservoir. It is noteworthy that $\delta W_{+}>0$ means that
the incoming scattering rate from the reservoir to the Rabi qubit exceeds
the outgoing rate from the qubit to the reservoir, $\tau _{R}>\tau _{0}$.
Hence, the Rabi oscillations, which in the pseudospin language are described
as a precession of a $\bm P_{\bot }$ around $z$ axis, decay the slower, the
higher {incoming scattering rate and, hence, the} occupation of the ground
state are. The population imbalance $P_{z}$ decays with a different time
constant $\tau _{rel}=\tau _{+}\left/ {}\right. \left( 1-\delta W_{+}\tau
_{+}\right) $.}

The decay rate of {Rabi oscillations can be recast in a different form in
order to demonstrate that $\tau _{R}>0$ despite of the negative sign in Eq.~(%
\ref{tau}). We introduce} the incoming and outcoming scattering rates as: $%
W_{1,2}^{\mathrm{in}}=\mathcal{W}_{1,2}^{\mathrm{in}}N_{R}$, $W_{1,2}^{%
\mathrm{out}}=\mathcal{W}_{1,2}^{\mathrm{out}}(1+N_{R})$, where $\mathcal{W}%
_{i}^{in/out}$ ($i=1,2$) are some constants and the occupation of the
reservoir $N_{R}$ must satisfy the following equation:

\begin{equation}
\dot{N}_{R}=P-\sum_{i=1,2}{\left[ \mathcal{W}_{i}^{\mathrm{in}}\left(
1+N_{i}\right) N_{R}-\mathcal{W}_{i}^{\mathrm{out}}N_{i}\left(
1+N_{R}\right) \right] }.  \label{NR}
\end{equation}%
Here $P$ is the particle generation rate in the reservoir. In what follows
we assume that the outscattering processes to the reservoir can be neglected
($\mathcal{W}_{1,2}^{\mathrm{out}}=0$) to simplify the subsequent
computations. Thus, in the steady-state, 
\begin{equation}
N_{i}\tau _{i}^{-1}-N_{i}N_{R}{\mathcal{W}}_{i}^{\mathrm{in}}=N_{R}{\mathcal{%
W}_{i}^{\mathrm{in}}}~~~i=1,2,  \label{N16}
\end{equation}%
and making use of Eqs.~(\ref{tau}), \eqref{N}, \eqref{NR} and the definition
of $\delta W_{+}$ we obtain 
\begin{equation}
{\tau _{R}^{-1}}={\tau _{\bot }^{-1}}+{N_{R}}\left( {\mathcal{W}}_{1}^{%
\mathrm{in}}N_{1}^{-1}+{\mathcal{W}}_{2}^{\mathrm{in}}N_{2}^{-1}\right)
\left/ 2\right. >0.  \label{tauR1}
\end{equation}%
It is clearly seen that an increase of the occupation of the doublet lease
to the decrease of $1/\tau _{R}$ and the increase of the coherence time of
Rabi oscillations. The closed form result for the $\tau _{R}$ dependence on
the pumping rate can be obtained taking into account that the steady-state
occupancy of the reservoir can be found from the following equation 
\begin{equation}
P=\frac{{\mathcal{W}}_{1}^{\mathrm{in}}N_{R}}{1-\tau _{1}{\mathcal{W}}_{1}^{%
\mathrm{in}}N_{R}}+\frac{{\mathcal{W}}_{2}^{\mathrm{in}}N_{R}}{1-\tau _{2}{%
\mathcal{W}}_{2}^{\mathrm{in}}N_{R}}.  \label{NR:oc}
\end{equation}%
Finally, we obtain 
\begin{multline}
{\tau _{R}}^{-1}={\tau _{0}}^{-1}-\frac{1+\alpha }{4\alpha }\frac{P\left(
\alpha \tau _{1}+\tau _{2}\right) +\alpha +1-}{P\tau _{1}\tau _{2}+\tau
_{2}+\tau _{1}}  \label{tau_R} \\
\frac{-\sqrt{P^{2}\left( \alpha \tau _{1}-\tau _{2}\right) ^{2}+2P\left(
\alpha \tau _{1}-\tau _{2}\right) \left( \alpha -1\right) +\left( \alpha
+1\right) ^{2}}}{P\tau _{1}\tau _{2}+\tau _{2}+\tau _{1}},
\end{multline}%
where $\alpha =\mathcal{W}_{1}^{\mathrm{in}}/\mathcal{W}_{2}^{\mathrm{in}}$
and it is assumed that $0\leq \alpha \leq 1$. For $\alpha>1$ in Eq.~%
\eqref{tau_R} one has to replace $\alpha\to 1/\alpha$ and exchange $\tau_1$
and $\tau_2$.

\begin{figure}[t]
\includegraphics[width=0.45\textwidth]{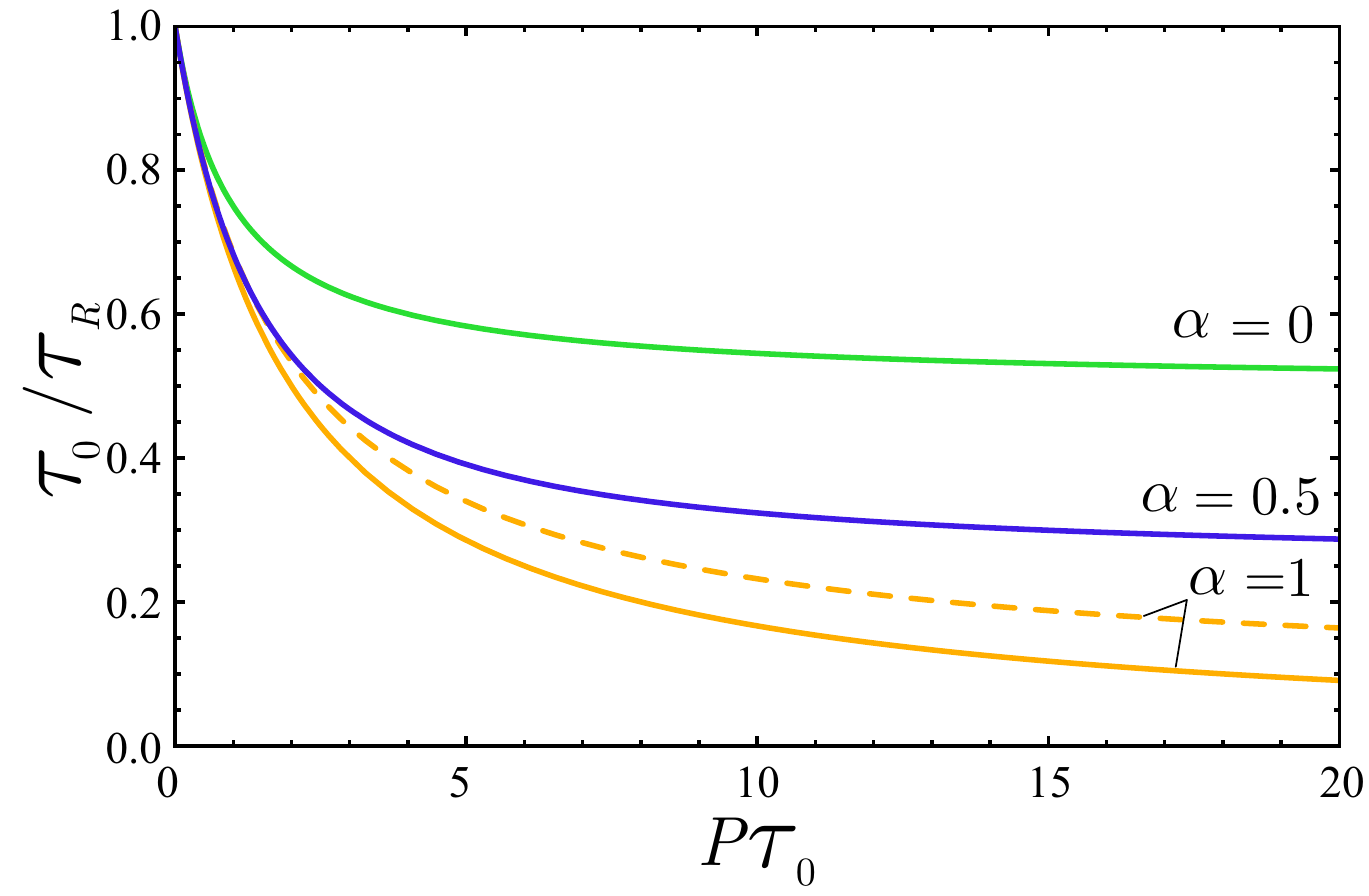}
\caption{(Color online) Ratio $\protect\tau_0/\protect\tau_R$ as a function
of the pumping rate $P\protect\tau_0$. For solid (dashed) curves $\protect%
\tau_1 = \protect\tau_2$ ($\protect\tau_2 = 10 \protect\tau_1$).}
\label{fig:t0tr}
\end{figure}

Figure \ref{fig:t0tr} shows the dependence of $\tau _{0}\left/ {%
\vphantom {t
\tau _{R} }}\right. \tau _{R}$ on $P\tau _{0}$ calculated for various values
of $\alpha $. {Here, similarly to Fig.~\ref{fig:Rabi} we consider} $%
\left\vert C_{x,p}\right\vert ^{2}=1/2$ that implies equal upper and lower
branch polariton lifetimes, i.e. $\tau _{pol}^{{}}\equiv \tau _{1}^{{}}=\tau
_{2}^{{}}$. Experimentally, this condition can be verified in specially
designed pillar microcavities \cite{Maragkou}, while in planar cavities the
lifetime of a $|UP\rangle $ state is usually much shorter than the lifetime
of an $|LP\rangle $ state. The imbalance of lifetimes in this case can be
compensated by the imbalance of pumping, which may be achieved in the case
of a quasi-resonant pumping of the $|UP\rangle $ state and can be accounted
for in our model by a proper choice of the parameter $\alpha $. We shall
also assume $\tau _{\perp }\gg \tau _{0}$. In this limit, the effective
lifetime of the polaritonic system $\tau _{0}$ approaches to $\tau
_{0}\simeq \tau _{pol}$. If, by contrast, $\tau _{\perp }$ is comparable
with $\tau _{pol}$ than at $P\tau _{pol}\rightarrow \infty $ 
\begin{equation}
\frac{\tau _{pol}}{\tau _{R}}\simeq \frac{\tau _{pol}}{\tau _{\bot }}+\frac{%
1-\alpha }{2}{,}  \label{rel_tau}
\end{equation}%
meaning that the decay time of Rabi oscillations is limited by $\tau _{\perp
}$. In any case, the decay rate of Rabi oscillations decreases with the
increase of the pumping rate, but the asymptotic value depends on the
imbalance of the scattering rates towards $\left\vert LP\right\rangle $ and $%
\left\vert UP\right\rangle $ states having its minimum at $\alpha =1$, i.e. $%
\mathcal{W}_{1}^{\mathrm{in}}=\mathcal{W}_{2}^{\mathrm{in}}$, cf.~%
\eqref{rel_tau} and Fig.~\ref{fig:t0tr}. In the optimum case, which would
correspond to a pillar microcavity with equal $|LP\rangle $ and $|UP\rangle $
lifetimes, $\alpha =1$, $\tau _{\perp }\rightarrow \infty $, for $\tau
_{pol}=10$~ps and ground state occupation $N=10^{2}$ we have $\tau _{R}\sim
1 $~ns.

In planar microcavities the lifetimes of $\left\vert UP\right\rangle $ and $%
\left\vert LP\right\rangle $ states can be strongly different. Importantly,
an increase of the coherence time $\tau _{R}$ can be observed in this case
as well: see dashed curve in Fig.~\ref{fig:t0tr}. Particularly, if $\mathcal{%
W}_{1}^{in}/\mathcal{W}_{2}^{in}=\tau _{2}/\tau _{1}$ the populations of
upper and lower polariton states become equal and $\tau _{R}=\tau
_{0}(1+P\tau _{0}/2)$.

Let us study the properties of the polariton qubit described by Eq.~%
\eqref{qubit}. Without any loss of generality one can rewrite Eq.~%
\eqref{qubit} as 
\begin{equation}
{\left\vert \Psi \right\rangle }=e^{-i\omega _{0}t}\left[ \cos \left( \frac{%
\theta }{2}\right) {\left\vert 0\right\rangle }+e^{i\pi /2}\sin \left( \frac{%
\theta }{2}\right) {\left\vert 1\right\rangle }\right] ,  \label{psi}
\end{equation}%
where we introduced the azimuthal angle $\theta =\Omega _{R}t$, states ${%
\left\vert 0\right\rangle }=\left( e^{-i\varphi }{\left\vert LP\right\rangle 
}+{\left\vert UP\right\rangle }\right) \left/ {}\right. \sqrt{2}$ and ${%
\left\vert 1\right\rangle }=\left( e^{-i\varphi }{\left\vert LP\right\rangle 
}-{\left\vert UP\right\rangle }\right) \left/ {}\right. \sqrt{2}$ represent
orthogonal (computational) qubit states with $\varphi $ being an arbitrary
phase, $\omega _{0}={\left( \Omega _{ph}+\Omega _{exc}\right) }\left/
{}\right. 2$, $\Omega _{ph}$ and $\Omega _{exc}$ being the bare frequencies
of the photon and exciton modes, respectively.

\begin{figure}[hptb]
\includegraphics[width=0.35\textwidth]{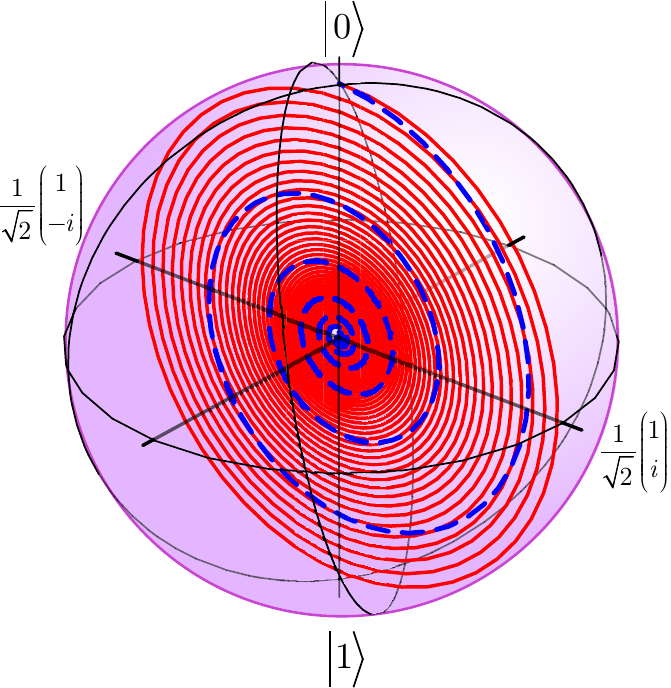}
\caption{(Color online) Bloch sphere representation of polariton qubit
dynamics without (dashed curve) and with (solid curve) reservoir supported
Rabi oscillations. The parameters are $\protect\alpha=1$, $\protect\varphi= 
\protect\pi \left/ \right.2$, $P \protect\tau_{0} = 20$. }
\label{fig:sphere}
\end{figure}

Figure~\ref{fig:sphere} shows the time evolution of the qubit state ${%
\left\vert \Psi \right\rangle }$ for various ratios of ${\tau _{0}%
\mathord{\left/ {\vphantom {\tau _{0}  \tau _{R} }} \right.
\kern-\nulldelimiterspace}\tau _{R}}$ at the Bloch sphere. Since one of the
Euler angles is equal to $\pi \left/ 2\right. $, see Eq.~\eqref{psi}, the
qubit Bloch vector evolves in a plane. In particular, it is clearly seen
that due to the reservoir supported Rabi oscillations the decoherence
effects are essentially suppressed for the solid (red) trajectory.

The manipulation of the qubit state Eq.~\eqref{psi} can be tailored through
variation of phase for a fixed Rabi frequency $\Omega _{R}$. On the other
hand, it is possible to manipulate by $\Omega _{R}$ as well by controlling
the exciton-photon coupling parameter $g$ by external electric fields, which
affect the exciton oscillator strength, or by manipulating the
exciton-photon detuning $\Delta $, cf.~\cite{arXiv:1312.2536}. In the latter
case, however, the Hopfield coefficients $C_{p}$ and $C_{x}$ also vary and
the states ${\left\vert 0\right\rangle }$ and ${\left\vert 1\right\rangle }$
are no more basic computational states of the system. Hence, it is
instructive rewrite Eq.~\eqref{psi} in the exciton-photon basis in this
case. In particular, from Eqs.~\eqref{doublet}, \eqref{psi} we express ${%
\left\vert \Psi \right\rangle }=e^{-i\omega _{0}t}\left[ \beta _{1}{%
\left\vert X\right\rangle }+\beta _{2}{\left\vert P\right\rangle }\right] ,$
where the coefficients $\beta _{1,2}$ are defined as $\beta _{1,2}=\left(
e^{-i\theta /2}C_{p,x}\pm e^{i\theta /2-i\varphi }C_{x,p}\right) \left/
{}\right. \sqrt{2}$, cf. Eq.~\eqref{qubit}. In this form the qubit state $%
\left\vert \Psi \right\rangle $ represents a linear superposition of matter
(excitonic) and photonic qubit state. The phase $\varphi $ determines the
initial state of the qubit. The short pumping pulse sets the initial
condition of $\beta _{2}=1$ that corresponds to the purely photonic state ${%
\left\vert \Psi (t=0)\right\rangle }={\left\vert P\right\rangle }$ .

Next, for quantum information applications, it is important to demonstrate
the entanglement between different qubit states. Such an entanglement could
be achieved with use of the coupled cavity architecture, cf.~\cite%
{PhysRevA.78.062336,PhysRevB.59.5082}. In this case, the entanglement can be
achieved due to the photonic tunneling between neighboring cavities.

Polarton qubits can be used for quantum cloning and quantum memory
applications \cite{EurPhysJD.58.1}. In particular, the quantum cloning
procedure of photonic state onto UP and LP states can be realized by using
an algorithm proposed by some of the present authors in \cite%
{J.Rus.Las.Res.27.482} which implies the coherent manipulation of Hopfield
coefficients.

Furthermore, the dynamical memory algorithm (see e.g. \cite%
{PhysRevA.86.013813}) can be realized using the semiconductor microcavity
structures described above. This algorithm imposes mapping of the quantum
information contained in the initially prepared photonic state ${\left\vert
\Psi (t=0)\right\rangle }={\left\vert P\right\rangle }$ onto the excitonic
qubit state ${\left\vert X\right\rangle }$. Writing, reading and storage
stages in this case can be achieved by a time control of the exciton dipole
matrix element [parameter $g\equiv g(t)$] and/or by manipulation of the
exciton-photon detuning $\Delta \equiv \Delta (t)$ adiabatically, cf.~\cite%
{arXiv:1312.2536}. We expect that the proposed mechanism of enhancement of
Rabi oscillations would allow realization of high temperature quantum
memories with a lifetime of the order of hundred of picoseconds.


\textit{Acknowledgments.} The financial support from the Russian Ministry of
Education and Science (Contract No. 11.G34.31.0067 with SPbSU and leading
scientist A. V. Kavokin), RFBR and EU projects POLAPHEN, SPANGL4Q and
LIMACONA is acknowledged. This work is also supported by the Russian
Ministry of Education and Science state task 2014/13. A. P. Alodjants
acknowledges support from ``Dynasty'' Foundation.


\end{document}